\documentclass[preprint,showpacs,preprintnumbers,amsmath,amssymb, showkeys,superscriptaddress,titlepage]{revtex4-2}
\usepackage[utf8]{inputenc}
\usepackage[T1]{fontenc}
\usepackage{amsmath, amssymb}
\usepackage{graphicx}
\usepackage{xcolor}
\usepackage[colorlinks=true, linkcolor=blue, citecolor=blue, urlcolor=blue]{hyperref}
\usepackage{url}

\begin{document}	
	
\title{Current problems of studying relativistic dissociation of light nuclei in nuclear emulsion}

\author{D. A. Artemenkov}
\affiliation{Joint Institute for Nuclear Research (JINR), Dubna, Russia}

\author{N.K. Kornegrutsa}
\affiliation{Joint Institute for Nuclear Research (JINR), Dubna, Russia}

\author{N. Marimuthu}
\affiliation{Joint Institute for Nuclear Research (JINR), Dubna, Russia}

\author{N.G. Peresadko}
\affiliation{P.N. Lebedev Physical Institute of the Russian Academy of Sciences (LPI), Moscow, Russia}

\author{V. V. Rusakova}
\affiliation{Joint Institute for Nuclear Research (JINR), Dubna, Russia}

\author{A.A. Zaitsev}
\affiliation{Joint Institute for Nuclear Research (JINR), Dubna, Russia}
\affiliation{P.N. Lebedev Physical Institute of the Russian Academy of Sciences (LPI), Moscow, Russia}
\email{zaicev@jinr.ru}

\author{P.I. Zarubin}
\affiliation{Joint Institute for Nuclear Research (JINR), Dubna, Russia}
\affiliation{P.N. Lebedev Physical Institute of the Russian Academy of Sciences (LPI), Moscow, Russia}

\author{I.G. Zarubina}
\affiliation{Joint Institute for Nuclear Research (JINR), Dubna, Russia}

\date{\today}
	
	\begin{abstract}
		The progress of the study of unstable states in relativistic dissociation events of light nuclei in nuclear emulsion is presented. Identification of these states is possible by means of the invariant mass determined from the most accurate and complete measurements of relativistic fragment emission angles in the approximation of conservation of momentum per nucleon of the parent nucleus. It is established that excitations $^{12}\mathrm{C}(0^{+}_{2})$ and $^{12}\mathrm{C}(3^{-})$ lead in the dissociation $^{12}\mathrm{C} \rightarrow 3\alpha$ and $^{16}\mathrm{O} \rightarrow 4\alpha$. The contribution of $^{9}\mathrm{B}$ and $^{12}\mathrm{C}(0^{+}_{2})$ decays to the leading channel of $^{3}\mathrm{HeH}$ dissociation of the $^{14}\mathrm{N}$ nucleus is estimated. The motivation and the beginning of the analysis of the relativistic dissociation $^{16}\mathrm{O}$$\rightarrow$$^{12}\mathrm{C}\alpha$ are presented. The presented relativistic dissociation events at the $^{7}\mathrm{Be}$$\rightarrow$$^{6}\mathrm{Li}p$ and $^{11}\mathrm{C}$$\rightarrow$$ ^{7}\mathrm{Be}\alpha$ coupling threshold point to the prospect of moving beyond $\alpha$-particle clustering.
	\end{abstract}
	
	\maketitle
	
\section{Introduction}
As a key aspect of the light nucleus structure and nucleosynthesis reactions, nucleon clustering remains the focus of experiments in the range of incident nuclei up to several tens of MeV per nucleon (reviews \cite{Freer-2014, Lombardo-2023}). At the same time, the region of limiting fragmentation of nuclei, which seems very remote from this topic, contains the potential for applying the established concepts of clustering and their extraordinary expansion. With the attainment of energies of several GeV per nucleon, the kinematic regions of fragmentation of colliding nuclei are clearly separated, and the duration of collisions is minimal. These factors can facilitate the interpretation of the final states of relativistic fragments that remain internally non-relativistic. However, it is difficult to take advantage of these advantages due to the radical reduction in ionization produced by fragments, as well as their magnetic rigidity, often poorly distinguishable from the primary beam.

Being observed with unique completeness and resolution, the tracks arising in the peripheral interactions of relativistic nuclei in nuclear emulsion demonstrate the generation of many-particle states of light, lightest nuclei and nucleons at the binding thresholds with probabilities determined by the structure of the parent nuclei (review \cite{Zarubin-2014}). Along with the collimation of relativistic fragments in a small solid angle determined by the Fermi motion of nucleons, there are no thresholds for their detection. Continuing with a recent publication and references therein \cite{Artemenkov-2024}, the progress of the BECQUEREL \cite{BECQUEREL} experiment in which the emulsion based relativistic approach is applied to systematical study phenomena related to the clustering in light nuclei is presented below.

The $^8$Be and $^9$B nuclei and a number of excitations of light isotopes that are unstable to the emission of $\alpha$-particles and nucleons at the binding thresholds have lifetimes of the order of femtoseconds or widths from eV to keV \cite{Ajzenberg-1988}. Unusually long-lived on the nuclear scale, they can be combined into a special class of unstable states arising at the lower limit of nuclear density and temperature. In the concepts of molecular-like or $\alpha$-condensate structures, they are represented as spatially separated groups of nucleons bound into clusters \cite{Funaki-2009,Wiringa-2000,Epelbaum-2011,Epelbaum-2012,Tohsaki-2017,Zhou-2020,Zhou-2023,Shen-2023}. Their occurrence in the final states of collisions of light nuclei may indicate the implementation of conditions corresponding to extremely low-energy nucleosynthesis reactions. Reconstructions of decays of known states of this type allow searching for analogs decaying into them. Exotic large sizes of unstable states allow one to propose a scenario for their formation in the resonant interaction of pairs of fragments with minimal values of invariant Lorentz factors of relative motion, and then the subsequent pickup of other fragments \cite{Zaitsev-2021}.

The ground state $^8$Be($0^+$) at 92 keV above the 2$\alpha$ threshold tops the list, followed by the Hoyle state of $^{12}$C($0^+_2$), decaying to $^8$Be($0^+$)$\alpha$ at 287 keV. Similar $^{16}$O levels are concentrated above at 1.7, 2.7, 3.8, and 4.4 MeV the threshold of $^{12}$C($0^+_1$)$\alpha$ (7.16 MeV) and below $^{12}$C($2^+_1$)$\alpha$ (4.5 MeV) and $^{15}$N$p$ (12.1 MeV). With $^9$B above the $^8$Be($0^+$)$p$ threshold (185 keV), the diversity is enriched in non-$\alpha$-fold states. In $^{11}$B there is a doublet of narrow levels above the $^7$Li$\alpha$ threshold (8.7 MeV) at 257 and 520 keV and below for $^{10}$Be$p$ (11.2 MeV). In $^{11}$C there are their analogs at 561 and 875 keV above the $^7$Be$\alpha$ threshold (7.6 MeV) and below $^{10}$B$p$ (8.7 MeV). The $^{10}$B nucleus has several narrow levels above the $^6$Li$\alpha$ threshold (4.5 MeV) and below $^9$Be$p$ (6.6 MeV). The absence of their analogs in the $^{10}$C nucleus above the $^6$Be$\alpha$ threshold (5.1 MeV) can be associated with the competition of $^9$B$p$ (4.0 MeV) and a high probability of $^8$Be($0^+$)$p$ fusion. As the excitation energy of nuclei goes beyond the limits of $\alpha$-clustering, narrow isobar-analog states (IAS) appear. This branch starts with the IAS of $^7$Be at 1.6 MeV above the threshold of $^6$Li$p$ (5.6 MeV). Its analogue in $^7$Li has not been established. The lifetime of the listed states allows one to consider them as participants in the dissociation channels of light nuclei considered below near the binding thresholds, which are essentially binary.

In nuclear emulsion layers longitudinally exposed to relativistic nuclei, the emission angles of relativistic fragments can be measured most accurately down to He and H. The best interpretation of the final states is possible in coherent dissociation events in which there are no tracks of target fragments (or ``white'' stars). For reference, video recordings of dissociation events taken with microscopes [movies] are collected (example in Fig. 1). Due to extremely low energy, decays of unstable states should appear as ensembles of fragments with the smallest opening angles and, consequently, minimal invariant masses. Thus, there are prerequisites for their uniform study. For ensembles arising near the corresponding coupling thresholds, the assumption that He fragments correspond to $\alpha$-particles, and H to protons, is justified. An accelerated search for the configurations of interest by transverse scanning of layers is possible.

\begin{figure}[]
	\centerline{\includegraphics[width=16cm]{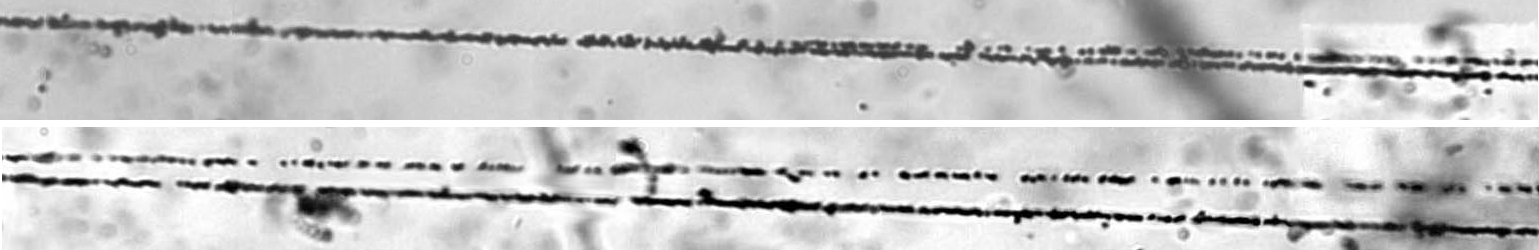}}
	\caption{Macrophotograph of the coherent dissociation event of the nucleus $^{16}$O $\to$ C + He with a momentum of 4.5 GeV/$c$ per nucleon; the grain size is no more than 0.5 $\mu$m. In the upper photograph on the left, the primary trace of $^{16}$O is visible, accompanied by short traces of $\delta$-electrons; the position of the peak is a sharp decrease in ionization. At a displacement of 1 mm in the direction of the fragment jet, their traces are distinguishable (lower photo).}
	\label{f1}
\end{figure}

The key decays of $^8$Be($0^+$), $^{12}$C($0^+_2$) and $^9$B are identified by upper limits on the invariant masses of pairs and triplets $Q_{2\alpha}$, $Q_{3\alpha}$ and $Q_{2\alpha p}$, determined in the approximation of conservation of momentum per nucleon of the parent nucleus \cite{Artemenkov-2020}. For convenience, the masses of the parent nucleus or groups of fragments are subtracted. In cases of incomplete inclusion of fragments in the mass calculation, a combinatorial background arises. However, it turns out to be insignificant in the initial parts of the distributions over this variable. In particular, the leading role of $^{12}$C($0^+_2$) and $^{12}$C($3^-$) decays in the relativistic dissociation of $^{12}$C nuclei was established in this way \cite{Artemenkov-2024}. Further justification of the approach used is provided by the presented estimate of the contribution of $^{16}$O $\rightarrow$ $^{12}$C($3^-$)$\alpha$ to 4$\alpha$-particle coherent dissociation.

The $^{12}$C($0^+_2$) enhancement appeared in the coherent dissociation $^{16}$O $\rightarrow$ 4$\alpha$ motivated a search in the distribution $Q_{4\alpha}$ for $^{16}$O($0^+_6$) excitation at only 660 keV above the 4$\alpha$ threshold (14.4 MeV) \cite{Artemenkov-2020}. Its relevance lies in testing the hypothesis of the structure of $^{16}$O($0^+_6$) as a 4$\alpha$-particle Bose-Einstein condensate decaying into $^{12}$C($0^+_2$)$\alpha$ or 2$^8$Be($0^+$) \cite{Tohsaki-2017}. The increase in the probability of $^8$Be($0^+$) and $^{12}$C($0^+_2$) with the number of $\alpha$-particles, subsequently established in the fragmentation of heavier nuclei, allows us to propose the fusion 2$\alpha$ $\rightarrow$ $^8$Be($0^+$)$\alpha$ $\rightarrow$ $^{12}$C($0^+_2$)$\alpha$ $\rightarrow$ $^{16}$O($0^+_6$) \cite{Artemenkov-2020}. At the same time, due to the extremely small energy gap of $^{16}$O($0^+_6$) $\rightarrow$ $^{12}$C($0^+_2$)$\alpha$, equal to only 296 keV, the theoretically estimated width of this decay is also extremely small, and the channel $^{16}$O($0^+_6$) $\rightarrow$ $^{12}$C($0^+_1$)$\alpha$ is proposed as an alternative \cite{Funaki-2009}, including in a framework of the search for decays of the predicted 5$\alpha$-condensate \cite{Zhou-2023}. The search for the decay $^{16}$O($0^+_6$) $\rightarrow$ $^{12}$C($0^+_1$)$\alpha$ is important in the context of nuclear astrophysical synthesis of the isotope $^{12}$C. It can serve as an alternative to the fusion $^8$Be($0^+$)$\alpha$ $\rightarrow$ $^{12}$C($0^+_2$) with the formation of $e^+e^-$ pairs or 2$\gamma$-decay $0^+$ $\rightarrow$ $2^+$ $\rightarrow$ $0^+$ with a probability of 1/2500. In the case of $^{16}$O($0^+_6$) $\rightarrow$ $^{12}$C($0^+_1$)$\alpha$, one of the $\alpha$-particles in the quartet serves as a kind of catalyst, removing the need for an electromagnetic transition. The coexistence of the decays $^{12}$C($0^+_2$)$\alpha$ and $^{12}$C($0^+_1$)$\alpha$ within the 165 keV width of $^{16}$O($0^+_6$) cannot be ruled out. The first results of the analysis of binary dissociation $^{16}$O $\rightarrow$ $^{12}$C($0^+_1$)$\alpha$ are highlighted below.

In the dissociation of isotopes of Be, B, C and N, including radioactive ones, the He and H ensembles are in the lead. For $^{10,11}$C and $^{10}$B, the decays $^9$B $\rightarrow$ $^8$Be($0^+$)$p$ \cite{Artemenkov-2020} and $^7$Be -- $^6$Be $\rightarrow$ $\alpha$2$p$ \cite{Kornegrutsa-2014} were identified. For $^{14}$N, indications were obtained of the contributions $^9$B $\rightarrow$ $^8$Be($0^+$)$p$ and $^{12}$C($0^+_2$) $\rightarrow$ $^8$Be($0^+$)$\alpha$ \cite{Artemenkov-2024}. In this context, the analysis of the dissociation of $^{14}$N presented below. In the dissociation of the non-$\alpha$-conjugate n isotopes, fragments heavier than $\alpha$-particles are formed. In a number of cases, the charge topology of such events indicates the mass numbers of the reaction participants, which allows one to calculate the invariant mass of the final ensemble. Then the search can be extended to states binding Li, Be, B, C with $\alpha$-particles. Its results are presented below for the very rare dissociation channels $^7$Be $\rightarrow$ $^6$Li$p$, $^{11}$C $\rightarrow$ $^7$Be$\alpha$.

\section{Coherent dissociation $^{16}$O $\rightarrow$ $^{12}$C($3^-$)$\alpha$}
In the distribution over $Q_{3\alpha}$ for the dissociation $^{12}$C $\rightarrow$ 3$\alpha$, in addition to $^{12}$C($0^+_2$), a peculiarity was observed between 2 and 4 MeV \cite{Artemenkov-2024}. Introducing the condition $^8$Be($0^+$) $Q_{2\alpha}$ $<$ 200 keV leads to a $Q_{3\alpha}$ doublet, as shown in Fig. 2 by a dotted line normalized to the number of events. The first peak with an average value of $Q_{3\alpha}$(RMS) = 417 $\pm$ 27 (165) keV corresponds to $^{12}$C($0^+_2$), and the second, described by the Rayleigh distribution with the parameter $\sigma$ = 2.4 $\pm$ 0.1 MeV -- to $^{12}$C($3^-$). For $^{12}$C($3^-$) the condition 1 MeV $<$ $Q_{3\alpha}$ $<$ 4 MeV is adopted. The effect of the $^8$Be($0^+$) condition can be associated with the suppression of the unidentified contribution of $^{12}$C $\rightarrow$ $^8$Be($2^+$)$\alpha$ and higher and broader excitations. Then the contributions of $^8$Be($0^+$), $^{12}$C($0^+_2$) and $^{12}$C($3^-$) to the dissociation of $^{12}$C $\rightarrow$ 3$\alpha$ are 43 $\pm$ 4, 9 $\pm$ 1, 19 $\pm$ 2\%, respectively. The contribution of $^{12}$C($0^+_2$) decays to the statistics of $^8$Be($0^+$) is 26 $\pm$ 4\%, and $^{12}$C($3^-$) is 45 $\pm$ 6\%. The ratio of $^{12}$C($0^+_2$) to $^{12}$C($3^-$) is 0.47 $\pm$ 0.06. While accounting for about a third of the statistics $^{12}$C $\rightarrow$ 3$\alpha$, in the case of $^{12}$C $\rightarrow$ $\alpha$$^8$Be($0^+$) the total contribution of $^{12}$C($0^+_2$) and $^{12}$C($3^-$) reaches two thirds.

Despite the increase in combinatory, $^8$Be($0^+$) and $^{12}$C($0^+_2$) appeared in the distributions of 641 coherent dissociation events $^{16}$O $\rightarrow$ 4$\alpha$ over $Q_{2\alpha}$ and $Q_{3\alpha}$ \cite{Artemenkov-2020}. When normalized to the number of events, an approximately twofold increase in $^8$Be($0^+$) and $^{12}$C($0^+_2$) was found compared to the $^{12}$C case. The main part of the distribution at $Q_{3\alpha}$ $>$ 1 MeV, extending to 20 MeV and having a wide maximum at 1 $<$ $Q_{2\alpha}$ $<$ 5 MeV, is described by the Rayleigh distribution with a parameter of 3.8 MeV. The coincidence of this value within the errors with the $^{12}$C case and the identification of $^{12}$C($3^-$) in it indicated the possibility of the presence of the $^{16}$O $\rightarrow$ $^{12}$C($3^-$)$\alpha$ channel.

The condition for the presence of $^8$Be($0^+$) in the event leads to the signal of $^{12}$C($3^-$), shown in Fig. 2. However, to estimate its contribution, it is desirable to weaken the combinatorial background in the $Q_{3\alpha}$ distribution. First of all, the already identified events can serve as a source, including 139 $^{12}$C($0^+_2$)$\alpha$ and 36 2$^8$Be($0^+$). After their removal, 196 events remain in the range adopted for $^{12}$C($3^-$), with an average value (RMS) of 2.48 $\pm$ 0.06 MeV (1.0). Among them, 105 have single such $\alpha$-triplets, and 91 have double ones, taken into account with a statistical weight of 0.5. The next contribution can be given by the events $^{16}$O $\rightarrow$ $^8$Be($0^+$)$^8$Be($2^+$), direct identification of which is impossible. Assuming the equality of $^8$Be($0^+$) and $^8$Be($2^+$) in the cases of $^{12}$C and $^{16}$O and their independent formation, the contribution of $^8$Be($0^+$)$^8$Be($2^+$) can be estimated as approximately equal to the contribution of 2$^8$Be($0^+$). It is subtracted from the number of $^{12}$C($3^-$) candidates. Taking this remark into account, it can be stated that $^{12}$C($3^-$) decays have been identified in the dissociation of $^{16}$O $\rightarrow$ 4$\alpha$. Then the contribution of the $^{12}$C($0^+_2$)$\alpha$ channel was 23 $\pm$ 2\%, $^{12}$C($3^-$)$\alpha$ -- 32 $\pm$ 2\%, 2$^8$Be($0^+$) -- 6 $\pm$ 1\%, and the ratio of the $^{12}$C($3^-$)$\alpha$ and $^{12}$C($0^+_2$)$\alpha$ channels was 1.4 $\pm$ 0.1.

\begin{figure}[]
	\centerline{\includegraphics[width=12cm]{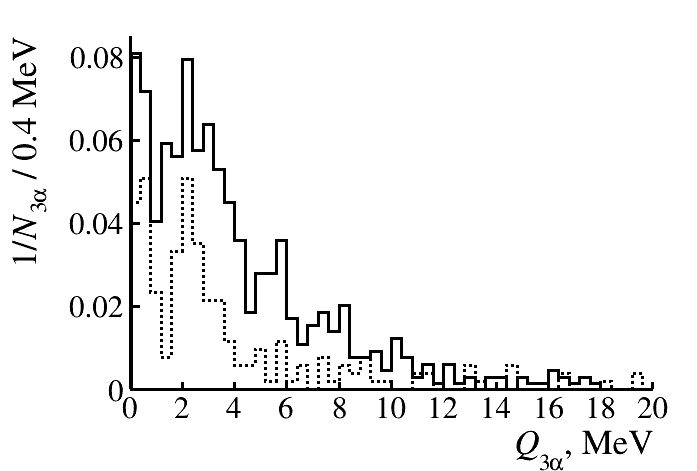}}
	\caption{Distribution over $Q_{3\alpha}$ in the events $^{12}$C $\to$ $^{8}$Be(0$^+$)$\alpha$ (dots) and $^{16}$O $\to$ $^8$Be(0$^+$)2$\alpha$ (solid); normalized to the number of events.}
	\label{f2}
\end{figure}

\section{Dissociation $^{14}$N $\rightarrow$ 3$\alpha p$}
Dissociation of the $^{14}$N nucleus via the leading channel 3$\alpha$ + H can serve as a common source of $^9$B and $^{12}$C($0^+_2$) \cite{Artemenkov-2024, Mitsova-2022}. In turn, they can arise in decays of more complex states, for example, IAS $^{13}$N$^*$(15.1). The statistics of measured events $^{14}$N $\rightarrow$ 3$\alpha$ + H has been brought to 128, including 29 "white" stars, and $^{14}$N $\rightarrow$ 3$\alpha$ with target fragments - 54.

Figure 3a shows the $Q_{2\alpha}$ distribution of $\alpha$-particle pairs. The average value of $\langle Q_{2\alpha} \rangle$ in 62 events containing $\alpha$-pairs with opening angles $\Theta_{2\alpha}$ $<$ 6 mrad is 114 $\pm$ 10 keV at RMS 92 keV. For 62 $\alpha$-pairs (53 events) in the region $Q_{2\alpha}$ $<$ 0.2 MeV $\langle Q_{2\alpha} \rangle$ is 76 $\pm$ 7 keV at RMS 61 keV. The contribution of $^8$Be($0^+$) events satisfying this condition to the dissociation $^{14}$N $\rightarrow$ 3$\alpha$ + H is 41 $\pm$ 6\%.

Figure 3b shows the $Q_{2\alpha p}$ distribution of 2$\alpha p$ triplets. In the region $Q_{2\alpha p}$ $<$ 0.5 MeV, adopted for $^{12}$C($0^+_2$), there are 30 events with $\langle Q_{2\alpha p} \rangle$ = 263 $\pm$ 20 keV at RMS 127 keV. They allow one to estimate the contribution of $^9$B as 23 $\pm$ 4\%. 53\% of Be($0^+$) decays are attributed to $^9$B decays. In the case of dissociation of $^{10}$B nuclei, such a contribution was equal to 39\% of the total number of events, and for "white" stars -- 50\%.

Figure 3c shows the $Q_{3\alpha}$ distribution of 3$\alpha$ triplets. With the $Q_{3\alpha}$ $<$ 0.7 MeV cutoff adopted for $^{12}$C($0^+_2$), 13 events with $\langle Q_{3\alpha} \rangle$ = 431 $\pm$ 35 keV at 125 keV RMS or 10\% of the statistics of this channel are distinguished. Although these parameters correspond to the expected $^{12}$C($0^+_2$) signal, these events coincide approximately half with the events with $^9$B candidates. The events of the 3He channel without H in the fragmentation cone are not discussed in the context of $^{12}$C($0^+_2$) due to their smaller statistics and the possible contribution of the excitation decays $^9$Be(1.67) $\rightarrow$ 2$\alpha$($n$) \cite{Artemenkov-2024}.

Figure 3d shows the $Q_{3\alpha p}$ distribution of 3$\alpha p$ quadruples, for which this problem is not significant. It shows the expected position of the IAS signal $^{13}$N$^*$(15.065). When it decays to $^9$B$\alpha$, $^{12}$C($0^+_2$)$p$ or $^{12}$C($3^-$)$p$, $^8$Be($0^+$) should be present. However, introducing the condition on the presence of $^8$Be($0^+$) in the event closes this possibility at the given level of statistics. The condition on $^9$B or $^{12}$C($0^+_2$) leads to a radical compression of this distribution with $\langle Q_{3\alpha p} \rangle$ equal to 2.5 $\pm$ 0.1 MeV at RMS 0.6 MeV.

\begin{figure}[]
	\centerline{\includegraphics[width=13cm]{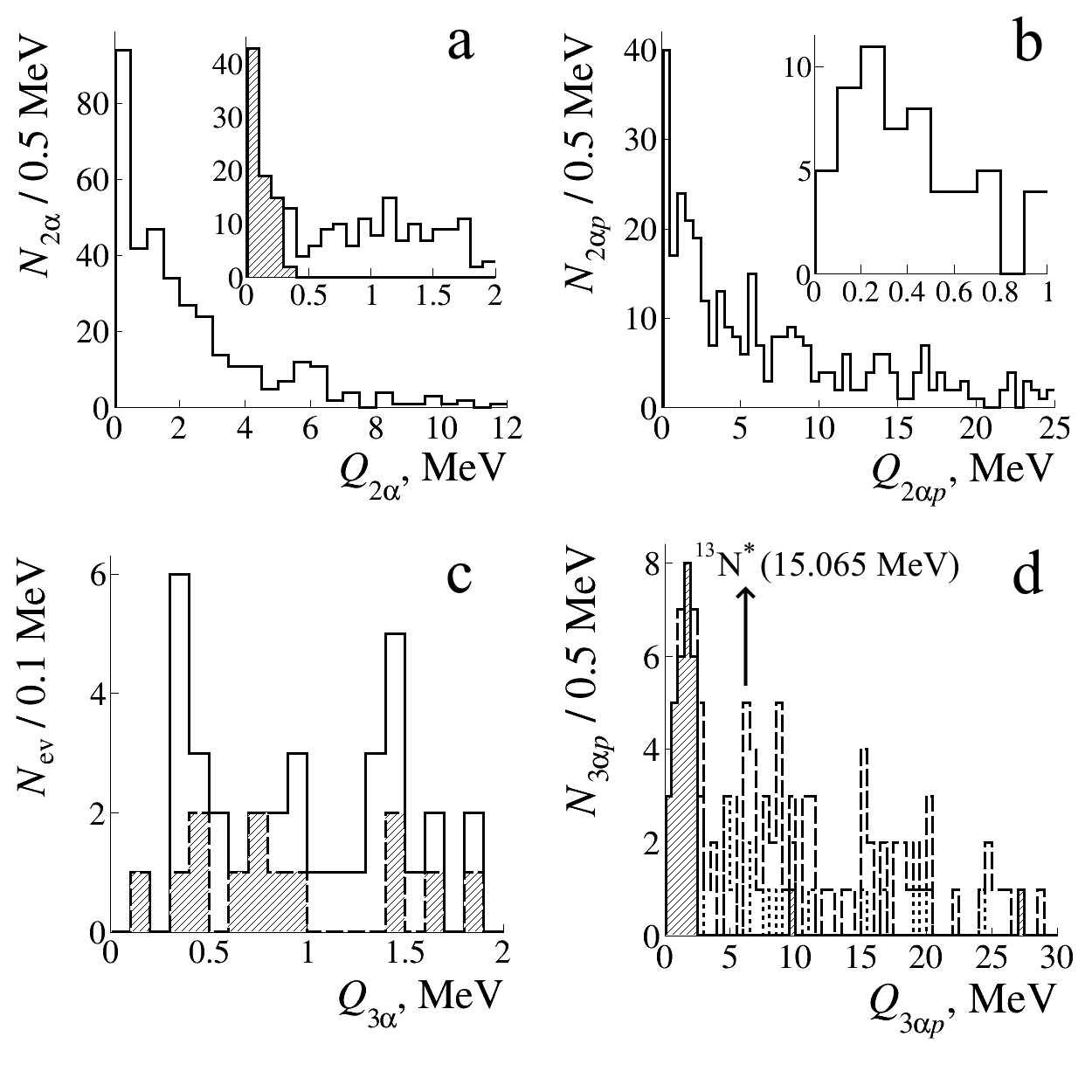}}
	\caption{Distribution over invariant masses $Q$ in the events $^{14}$N $\to$ 3$\alpha$ + H; a - $Q_{2\alpha}$ (hatching - $\Theta_{2\alpha}$ $<$ 6 mrad); b - $Q_{2\alpha p}$ (inset: $Q_{2\alpha p}$ $<$ 1 MeV); c - $Q_{3\alpha}$ (hatching - veto $^9$B); d - $Q_{3\alpha p}$, (dotted line - all events, dots - condition $^8$Be(0$^+$), hatching - condition $^9$B or $^{12}$C(0$^+_2$)).}
	\label{f3}
\end{figure}

\section{Dissociation $^{16}$O $\rightarrow$ $^{12}$C$\alpha$}
There are measurements of the emission angles and total momenta of fragments $P_{\rm fr}$ 11000 interactions of $^{16}$O nuclei at the momentum $P_0$ = 3.25 GeV/$c$ per nucleon in the 1-meter liquid hydrogen bubble chamber of JINR (VPK-100), placed in a magnetic field \cite{Glagolev-2001}. The $P_{\rm fr}$/$P_0$ ratios allow one to determine the isotopic composition of the fragments. To determine the charges of fragments heavier than He, it is sufficient to ensure that the sum of the charges of secondary particles in the event is equal to 9. The VPK-100 data allow one to check simplifying assumptions when interpreting observations in a nuclear emulsion. In particular, this was used to confirm the dominance of $^4$He in extremely narrow He pairs from $^8$Be($0^+$) decays. At the same time, due to insufficient resolution in the distribution of invariant masses of $Q_{2\alpha}$ pairs calculated from the measured $P_{\rm He}$ momenta, the $^8$Be signal practically disappeared. In contrast, the calculation of $Q_{2\alpha}$ with fixed 4$P_0$ momenta, depending only on the fragment emission angles measured in VPK-100, shows a peak of $^8$Be($0^+$) \cite{Zaitsev-2021-Ench}.

The assumption about the dominance of the $^{12}$C($0^+_1$)$\alpha$ pair in the C + He topology was verified on 214 3-way C + He (+$p$) events. Figure 4a shows the distribution over $P_{\rm He}$/$P_0$ and $P_{\rm C}$/$P_0$. The contribution of the $^{13}$C + $^3$He events with a threshold of 22.8 MeV is negligible. The $P_{\rm C}$/$P_0$ projection does not show signals of lighter C isotopes (fig. 4b), and the $P_{\rm He}$/$P_0$ projection does not show signals of $^3$He (fig. 4c), the formation of which would correspond to channels with thresholds above 30 MeV. The excitation of $^{12}$C($2^+_1$) can contribute to the broadening of the $P_{\rm C}$/$P_0$ distribution. Thus, the $^{12}$C + $\alpha$ channel dominate, which allows one to neglect the others. Figure 4d shows the distribution over the invariant mass of the $Q_{^{12}\rm{C}\alpha}$ pairs for the entire sample and 64 events highlighted in Fig. 4a. The calculations are performed in the approximation of conservation of the initial momentum per nucleon $P_0$ by fragments. It contains events both near threshold resonances and in the $^{16}$O($0^+_6$) region.

\begin{figure}[]
	\centerline{\includegraphics[width=13cm]{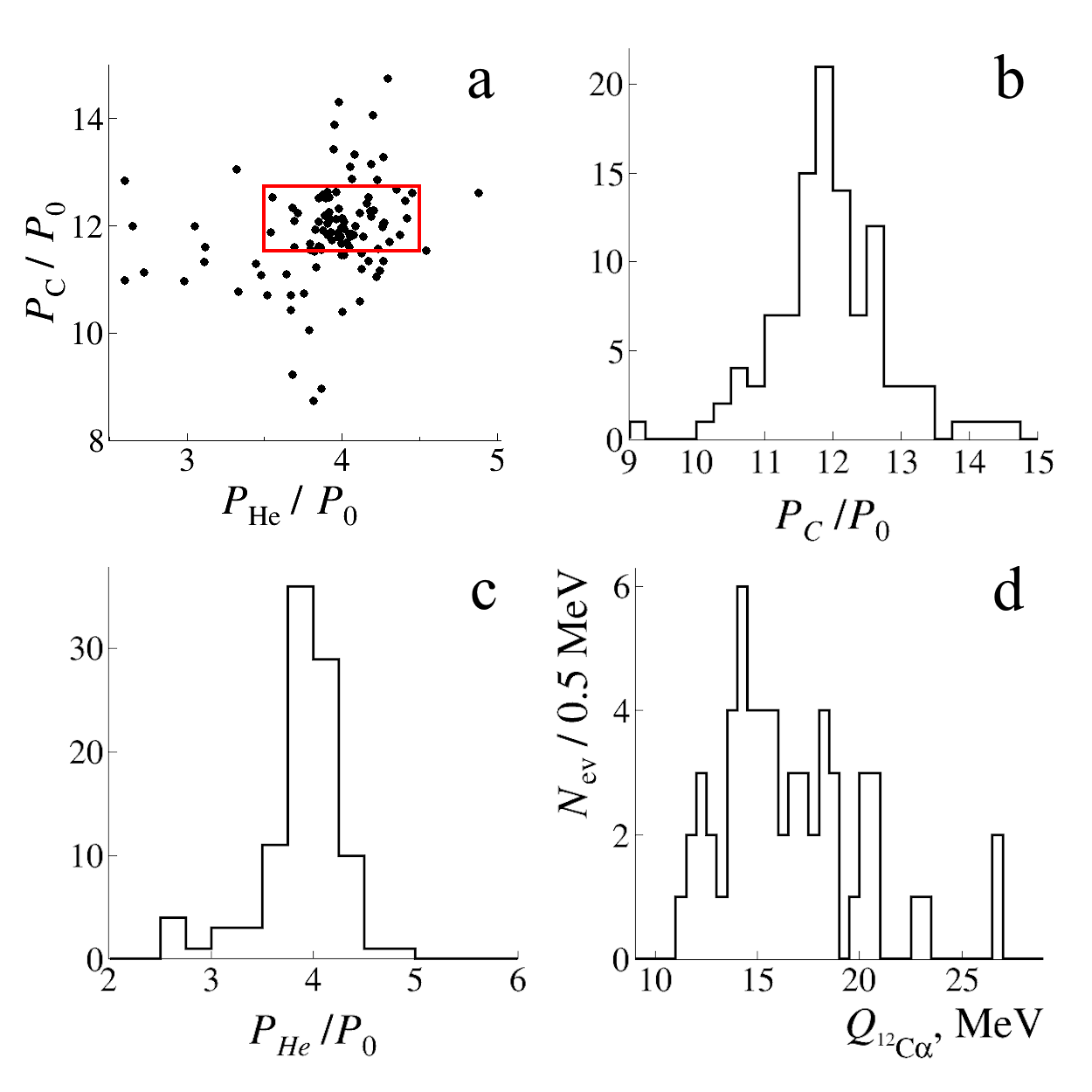}}
	\caption{Distribution of 3-prong events C + He (+$p$) in a hydrogen bubble chamber over $P_\mathrm{He}$/$P_{0}$ and $P_\mathrm{C}$/$P_{0}$ (a), projections (b and c) and invariant mass $Q_{^{12}\mathrm{C}\alpha}$ (d).}
	\label{f4}
\end{figure}

Recently, an accelerated search for C + He events has been started by transverse scanning of NTE layers irradiated in the 1980s at the JINR Synchrophasotron in a beam of $^{16}$O nuclei at 4.5 GeV/$c$ per nucleon. The directions in pairs of converging traces of He and a heavier fragment are measured and traced to the interaction vertex (Fig. 1). To date, 67 events have been measured, including 22 ``white'' stars. Distributions of polar emission angles $\theta_{\rm C}$ and $\theta_{\rm He}$ (Fig. 5a and b) were obtained from measurements of coordinates on secondary tracks. They are described by the Rayleigh distribution with parameters $\sigma_{\theta_{\rm C}}$ = (3.9 $\pm$ 0.3) mrad and $\sigma_{\theta_{\rm{He}}}$ = (8.4 $\pm$ 0.8) mrad, and by the opening angle $\Theta_{\rm CHe}$ $\sigma_{\alpha}$ = (9.4 $\pm$ 0.9) mrad (Fig. 5c).

\begin{figure}[]
	\centerline{\includegraphics[width=15cm]{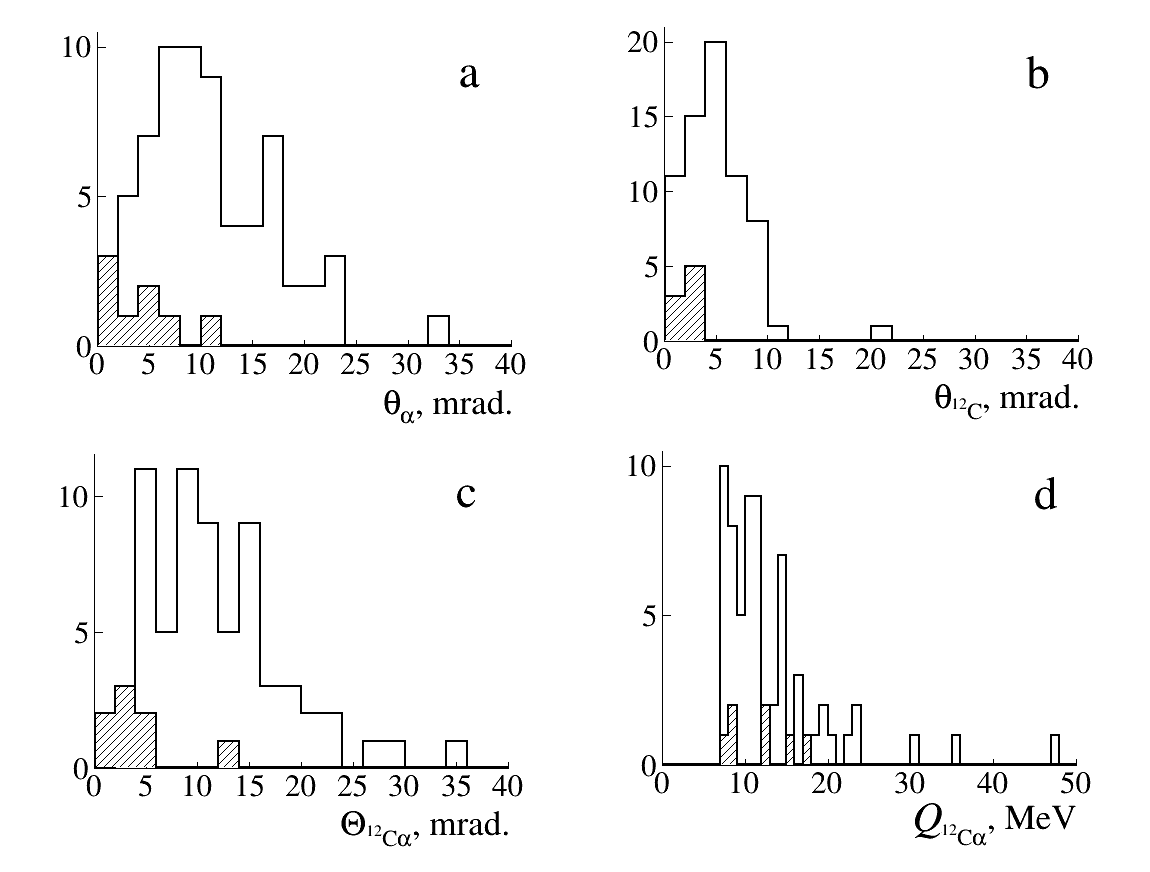}}
	\caption{Results of measurements of events $^{16}$O $\to$ $^{12}$C$\alpha$ at 4.5 and 15 GeV/$c$ (shaded): emission angles $\theta_{\alpha}$ (a) and $^{12}$C (b), scattering angles between them (c) and the invariant masses $Q_{^{12}\mathrm{C}\alpha}$.}
	\label{f5}
\end{figure}

These measurements allow one to obtain the distributions of the invariant mass of $Q_{^{12}\rm{C}\alpha}$ pairs under the assumption of conservation of momentum per nucleon and the $^{12}$C + $\alpha$ correspondence (Fig. 5d), which has a similar form to that shown in Fig. 6. Eight events, including two "white" stars, with an average value of $\langle Q_{^{12}\rm{C}\alpha} \rangle$ = 15.1 $\pm$ 0.2 MeV at RMS 0.6 MeV, can be classified as candidates for the decay of $^{16}$O($0^+_6$). This decay variant has the prospect of being confirmed as statistics grows.

In general, the obtained results stimulate further application of the invariant approach in the study of near-threshold unstable states arising from the relativistic dissociation of light nuclei in a nuclear emulsion.

\begin{figure}[]
	\centerline{\includegraphics[width=15cm]{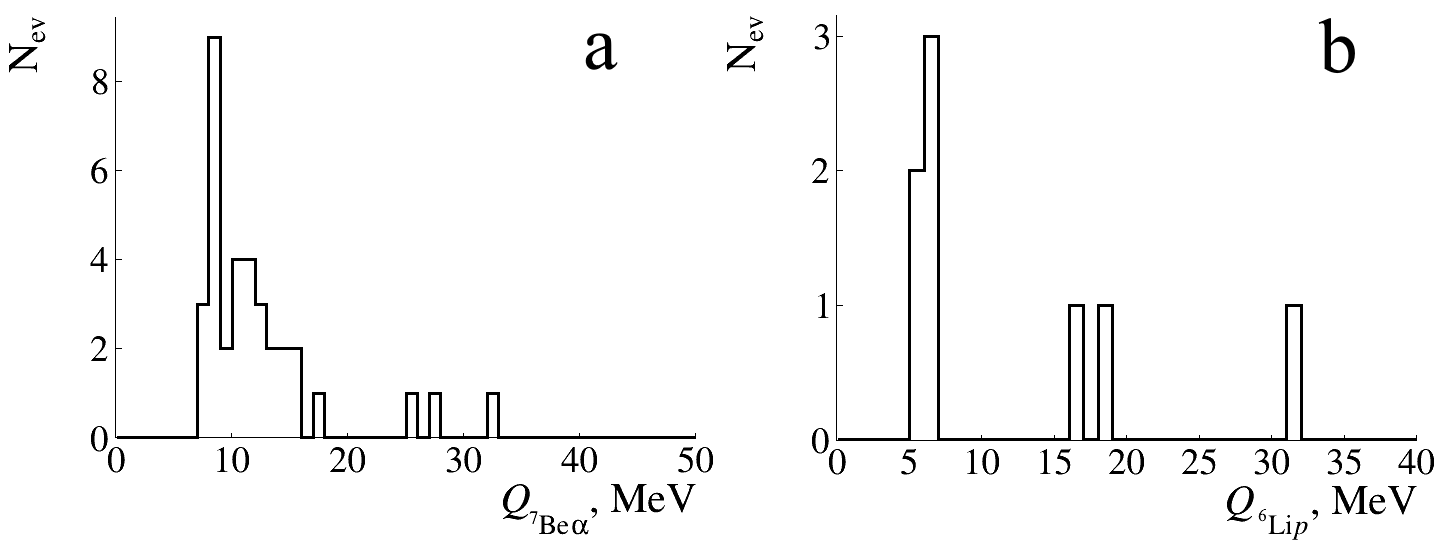}}
	\caption{Distribution of invariant masses of rare events $^{11}$C $\to$ $^7$Be$\alpha$ (a) and $^7$Be $\to$ $^6$Li$p$ (b).}
	\label{f6}
\end{figure}

\section{Rare events $^7$B$\mathrm{e}$ $\rightarrow$ $^6$L$\mathrm{i}$$p$ and $^{11}$C $\rightarrow$ $^7$B$\mathrm{e}$$\alpha$}

Along with the dominance of channels containing only He and H, rare events with the formation of a heavier fragment were observed in the relativistic dissociation of $^7$Be \cite{Peresadko-2007,Kornegrutsa-2014} and $^{11}$C \cite{Artemenkov-2015}. Their charge topology allows the identification of mass numbers. The invariant mass distribution $Q_{^6\rm{Li}p}$ indicates 8 events $^7$Be $\rightarrow$ $^6$Li$p$ above the 5.6 MeV threshold (Fig. 6). Their average value of 6.4 $\pm$ 0.3 MeV at RMS 0.6 corresponds to the $^7$Be(7.2) level with the width $\Gamma$ = 0.40 keV \cite{Ajzenberg-1988}. Despite the absence of isospin prohibition, the small width of this level is due to its difference from the main $^4$He-$^3$He configuration. The invariant mass distribution $Q_{^7\mathrm{Be}p}$ indicates 30 events $^{11}$C $\rightarrow$ $^7$Be$\alpha$ above the 7.6 MeV threshold (Fig. 6). 12 pairs at $Q_{^7\mathrm{Be}p}$ $<$ 9 MeV can be associated with a quartet of narrow $^{11}$C$^*$ levels (8.11, 8.42, 8.66, 8.70).

\section{Conclusion}
The contributions of the $^{12}$C($0^+_2$) and $^{12}$C($3^-$) decays, which constitute 11 and 19\% in the $^{12}$C $\rightarrow$ 3$\alpha$ dissociation, increase to 22 and 32\% in the $^{16}$O $\rightarrow$ 4$\alpha$ dissociation. They correspond to 26 and 45\% in $^{12}$C $\rightarrow$ $^8$Be($0^+$)$\alpha$ (43\%) and 37 and 52\% in $^{16}$O $\rightarrow$ $^8$Be($0^+$)2$\alpha$ (62\%). Thus, the first two excitations of $^{12}$C above the $\alpha$-particle binding threshold lead in the relativistic dissociation of these nuclei.

The number of events with target fragments in the $^{14}$N $\rightarrow$ 3He + H channel is twice the statistics of the $^{14}$N $\rightarrow$ 3He channel, coinciding with the $^{10}$B $\rightarrow$ 2He (+H) case, indicating a broader spatial distribution of neutrons compared to protons. This observation deserves to be applied to the fragmentation of neutron-rich nuclei, starting from $^{11}$B, to establish a neutron halo in them.

The contributions of $^9$B and $^{12}$C($0^+_2$) to the leading channel of 3HeH dissociation of the $^{14}$N nucleus are estimated at 23 and 10\%. However, there is an overlap. It is unclear whether the resolution limit has been reached or whether there is a more fundamental overlap of $^9$B and $^{12}$C($0^+_2$). The joint condition on $^9$B or $^{12}$C($0^+_2$) indicates that the $\alpha$-particles and protons in events with $^9$B or $^{12}$C($0^+_2$), but not attributed to decays, are close to them in the invariant mass 3$\alpha p$.

An analysis of the binary dissociation $^{16}$O $\rightarrow$ $^{12}$C($0^+_1$)$\alpha$ at 4.5 GeV/$c$ per nucleon has been started to search for the decay of the excited state $^{16}$O($0^+_6$) at 15.1 MeV. About two thirds of the 70 measured events have an invariant mass below 14 MeV, and 12\% above this threshold indicate such a possibility. The assumptions made about $^{12}$C($0^+_1$) and $\alpha$ were verified against hydrogen bubble chamber data in a magnetic field. Having a similar shape, the invariant mass distribution of such pairs confirms the possibility of identifying $^{16}$O($0^+_6$) with a multiple increase in statistics.

This conclusion stimulated the beginning of a similar analysis of the emulsion irradiated with $^{16}$O $\rightarrow$ nuclei at 15 GeV/$c$ per nucleon at AGS BNL in the 1990s. In this case, there are methodological advantages associated with greater beam compactness, a smaller contribution of associated $\alpha$-particles and a 3-fold smaller fragmentation cone. Invariant representation of the data from both irradiations will allow independent comparison of the conclusions and verification of the universality of $^{16}$O($0^+_6$) formation. The measurements of the first events found are added to Fig. 5 (shaded).

Rare events $^7$Be $\rightarrow$ $^6$Li$p$ and $^{11}$C $\rightarrow$ $^7$Be$\alpha$, concentrated at the thresholds of invariant masses, indicate the prospect of expanding the list of unstable states formed in relativistic dissociation. With angular resolution, the difference in magnetic rigidities allows investigating the dissociation channels $^{11}$C $\rightarrow$ $^7$Be$\alpha$ and $^{11}$B $\rightarrow$ $^7$Li$\alpha$ in electron experiments. The search for narrow mirror states $^{11}$B$^*$(8.92, 9.19) above the 8.7 MeV threshold of the $^7$Li$\alpha$ channel is continued within the framework of transverse scanning of layers irradiated in the primary $^{11}$B beam. An example of dissociation $^{11}$B event is shown in Fig. 7.

\begin{figure}[]
	\centerline{\includegraphics[width=16cm]{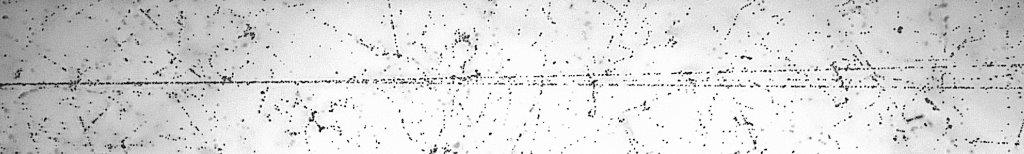}}
	\caption{Coherent dissociation of $^{11}$B $\to$ 2He + H was captured in nuclear track emulsion using an Olympus BX63 microscope at 60$\times$ magnification. The interaction vertex and He and H fragment tracks, emerging within a narrow angular cone, are clearly visible in the micro-photograph. The micro-photograph obtained by superimposing a series of frames from different focal planes made it possible to create a completely sharp photograph of an area 20 microns deep.}
	\label{f7}
\end{figure}
	
	\bibliographystyle{unsrt}

\end{document}